\documentstyle[times,pramana,epsf,floats]{ias}
\begin{document}
\mark{{High energy cosmic rays}{E. waxman}}
\title{High energy cosmic-rays: puzzles, models, and giga-ton neutrino telescopes}

\author{E. Waxman}
\address{Physics Faculty, Weizmann Institute, Rehovot, Israel}
\abstract{
The existence of cosmic rays of energies exceeding $10^{20}$~eV is one of
the mysteries of high energy astrophysics. The spectrum and the
high energy to which it extends rule out almost all suggested source
models. The challenges posed by observations to models for the origin of high energy cosmic rays are reviewed, and the implications of recent new experimental results are discussed. Large area high energy cosmic ray detectors and large volume high energy neutrino detectors currently under construction
may resolve the high energy cosmic ray puzzle, and shed light on the
identity and physics of the most powerful accelerators in the universe.}

\maketitle

\section{Introduction}

\begin{figure}[htbp]
\epsfxsize=12cm
\centerline{\epsfbox{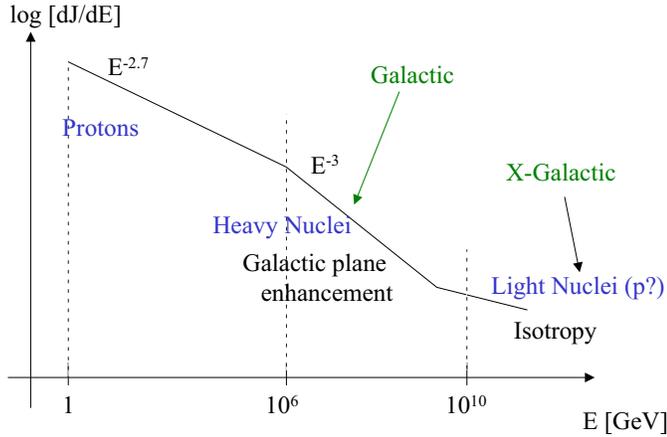}}
\caption{A schematic description of the differential spectrum and of the composition of cosmic rays observed at Earth.}
\label{fig:CRspec}
\end{figure}

Fig.~\ref{fig:CRspec} presents a schematic description of the spectrum and composition of cosmic rays observed at Earth \cite{CR_data_rev}. At low energies, $\sim1$~GeV per particle, the flux is dominated by protons. The average particle mass increases with energy and the composition becomes dominated by heavy nuclei at $\sim10^6$~GeV per particle, where the spectrum also becomes steeper. At still higher energy, $\sim10^{10}$~GeV, the spectrum and composition change again: the spectrum becomes harder (flatter) and there is strong evidence that the composition becomes lighter, likley dominated by protons. It therefore appears that a new source of cosmic rays comes to dominate above $10^{19}$~eV. Since heavy nuclei of energies $\lesssim10^{18}$~eV are confined by the Galactic magnetic field, it is believed that cosmic rays of energy $<10^{19}$~eV are of Galactic origin. This view is supported by the (small but statistically significant) enhancement, observed below $10^{19}$~eV, of cosmic ray flux from the direction of the Galactic plane. Protons of energy $>10^{19}$~eV are not confined by the Galactic magnetic field, and the isotropic distribution of cosmic rays above $10^{19}$~eV therefore suggests that the flux is dominated at these energies by an extra-Galactic source of protons.

The origin of the highest energy, $>10^{19}$~eV, cosmic rays (UHECRs) is a mystery. As explained in \S\ref{sec:acceleration}, the high energies observed rule out almost all candidate sources. The situation is further complicated by the interaction of high energy protons with microwave background photons. As explained in \S\ref{sec:GZK}, this interaction limits the propagation of protons of energy $>10^{20}$~eV to 
$\lesssim100$~Mpc, and there are no exceptionally bright sources that may be suspected as UHECR sources within such a distance from Earth. More over, it is not clear whether or not the expected "GZK suppression" \cite{gzk} of UHECR flux above $\sim5\times10^{19}$~eV, due to interaction with microwave background photons, is observed. These difficulties have led to the suggestion that modifications of the basic laws of physics are required in order to account for the existence of UHECRs. Such suggestions are discussed in detail in a separate contribution to these proceedings \cite{Drees}. The present contribution focuses on the discussion of possible solutions to the UHECR puzzle, which do not invoke modifications to the basic laws of physics. In \S\ref{sec:GRB} we discuss the gamma-ray burst (GRB) model for UHECR production and some of its predictions, which may be tested with giga-ton neutrino telescopes. A more detailed discussion of the model and its predictions for planned UHECR and neutrino detectors may be found in \cite{GRBCRrev}. Some general comments on the role that neutrino telescopes may play in resolving the UHECR puzzle are given in \S\ref{sec:GTnu}. Our main conclusions are summarized in \S\ref{sec:discussion}.

\section{Acceleration: General considerations and candidate sources}
\label{sec:acceleration}

\begin{figure}[htbp]
\epsfxsize=10cm
\centerline{\epsfbox{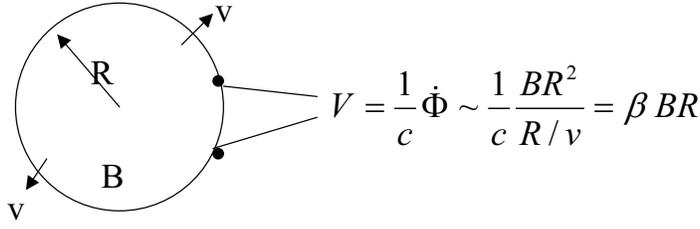}}
\caption{Potential drop generated by an outflow of magnetized plasma.}
\label{fig:acceleration}
\end{figure}

A detailed discussion of particle acceleration is beyond scope of this talk. However, the essence of the challenge can be understood using the following simple argument. Consider an astrophysical source driving a flow of magnetized plasma, with characteristic magnetic field strength $B$ and velocity $v$. Imagine now a conducting wire encircling the source at radius R, as illustrated in fig.~\ref{fig:acceleration}. The potential generated by the moving plasma is given by the time derivative of the magnetic flux $\Phi$ and is therefore given by $V=\beta B R$ where $\beta=v/c$. A proton which is allowed to be accelerated by this potential drop would reach energy
$E_p\sim\beta eB R$. The situation is somewhat more complicate in the case of a relativistic outflow, where $\Gamma\equiv(1-\beta^2)^{-1/2}\gg1$. In this case, the proton is allowed to be accelerated only over a fraction of the radius $R$, comparable to $R/\Gamma$. To see this, one must realize that as the plasma expands, its magnetic field decreases, so the time available for acceleration corresponds, say, to the time of expansion from $R$ to $2R$. In the observer frame this time is $R/c$, while in the plasma rest frame it is $R/\Gamma c$. Thus, a proton moving with the magnetized plasma can be accelerated over a transverse distance $\sim R/\Gamma$. This sets a lower limit to the product of the magnetic field and source size, which is required to allow acceleration to $E_p$,
\begin{equation}\label{eq:BR}
    BR>\Gamma E_p/e\beta.
\end{equation}

\begin{figure}[htbp]
\epsfxsize=10cm
\centerline{\epsfbox{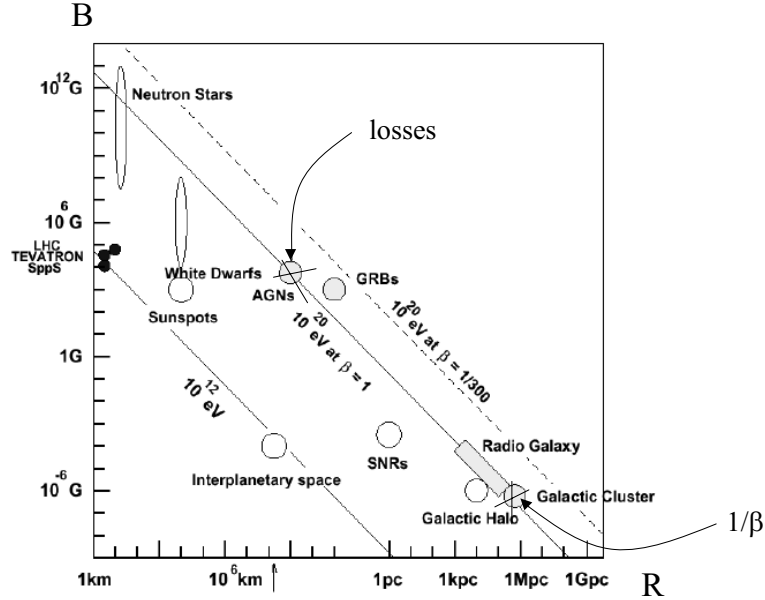}}
\caption{"Hillas plot" [5]. The magnetic field strength and size of different objects. In order to allow proton acceleration to $E_p$, sources must satisfy eq.~\ref{eq:BR}.}
\label{fig:Hillas}
\end{figure}
Figure~\ref{fig:Hillas} shows $B$ and $R$ for various objects. GRBs are the only type of sources which safely satisfies the basic requirement of eq.~\ref{eq:BR}. Active galactic nuclei (AGN) appear in two places: as "AGNs", where acceleration is considered near the central source (massive black hole) which drives the outflow, and as "Radio Galaxies", where acceleration is considered in regions of the flow far away from the central object, where interaction with surrounding gas leads to radio emission. Acceleration to high energy near the central source is impossible due to the high density of photons, which leads to rapid energy loss of protons by photo-production of pions. Acceleration in galaxy clusters is impossible since the flows there are characterized by $\beta\le10^{-2.5}$.

Eq.~\ref{eq:BR} also sets a lower limit to the rate $L$ at which energy should be generated by the source. The magnetic field carries with it an energy density $B^2/8\pi$, and the flow therefore carries with it an energy flux $>vB^2/8\pi$ (some energy is carried also as plasma kinetic energy), which implies $L>vR^2B^2$. Using eq.~\ref{eq:BR} we find
\begin{equation}\label{eq:L}
  L>\frac{\Gamma^2}{\beta}\left(\frac{E_p}{e}\right)^2c
  =10^{45.5}\frac{\Gamma^2}{\beta}\left(\frac{E_p}{10^{20}\rm eV}\right)^2{\rm erg/s}.
\end{equation}
Only two types of sources are known to satisfy this requirement. The brightest steady sources are active galactic nuclei (AGN). For them $\Gamma$ is typically between 3 and 10, implying $L>10^{47}{\rm erg/s}$. This is marginally satisfied by the brightest AGN. The brightest transient sources are GRBs. For these sources $\Gamma\simeq10^{2.5}$ implying $L>10^{50.5}{\rm erg/s}$, which is satisfied since the typical observed MeV-photon luminosity of these sources is $L_\gamma\sim10^{52}{\rm erg/s}$.

It was recognized early on (\cite{Hillas} and references therein) that while highly magnetized neutron stars may lie above the required line in the "Hillas plot" (fig.~\ref{fig:Hillas}), it is hard to utilize the potential drop in their electro-magnetic winds for proton acceleration to ultra-high energy. The mechanisms recently proposed for acceleration in "magnetars" (\cite{Arons} and references therein) also face serious difficulties. In \cite{Arons}, for example, the electro-magnetic wind must penetrate through a supernova envelope shell without losing energy to acceleration of the shell and without "contaminating" the wind with baryons. The mechanism by which such penetration may be achieved is unclear.

\section{Propagation: The GZK suppression}
\label{sec:GZK}

\subsection{Interaction with microwave background photons}
\label{sec:CMB}

\begin{figure}[htbp]
\epsfxsize=10cm
\centerline{\epsfbox{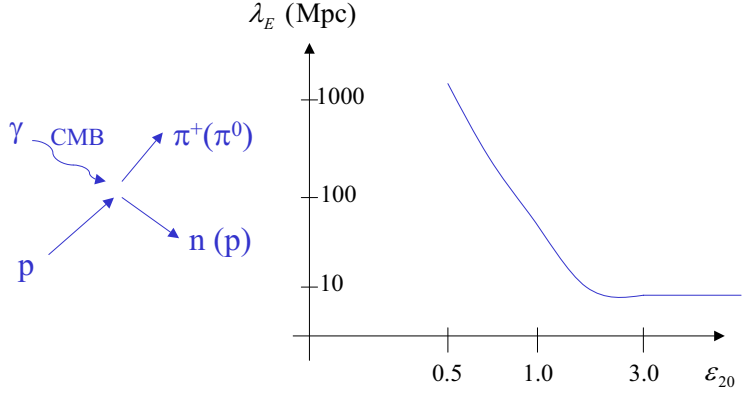}}
\caption{Schematic description of the strong reduction of the energy loss distance, $\lambda_E$, of high energy protons, due to pion production interactions with microwave background photons. $E_{20}$ is the proton energy in units of $10^{20}$~eV.}
\label{fig:gzk}
\end{figure}

As illustrated in Fig.~\ref{fig:gzk}, high energy protons may interact with cosmic microwave background photons to produce pions. In each interaction of this type, the proton loses a fraction $\sim m_\pi/m_p$ of its energy. The threshold energy requirement, $E_pE_\gamma\gtrsim m_p m_\pi c^4$ where $E_p$ and $E_\gamma$ are the proton and photon energies respectively, implies that protons of energy $E_p>10^{20}$~eV may interact with almost all of the $T=2.7^o$~K background photons, while protons of lower energy may interact only with the tail of the Planck distribution. Thus, the energy loss distance, $\lambda_E(E_p)$, drops rapidly with energy in the range of $0.5\times10^{20}$~eV to $3\times10^{20}$~eV (see Fig.~\ref{fig:gzk}).

The rapid drop of $\lambda_E(E_p)$ with increasing $E_p$ has two major consequences. First, it effectively limits the distance out to which sources may contribute to the UHECR flux above $10^{20}$~eV to less than $\sim100$~Mpc, a small distance on cosmological scale. In particular, there is no AGN bright enough to satisfy the constraint of Eq.~\ref{eq:L} with this distance. This appears to rule AGN as the sources of UHECRs. This difficulty may be avoided by assuming the existence of "dead AGN" \cite{Boldt}, massive black holes in nearby galaxies that produce UHECRs without producing almost any light. The difficulty here is that even if a small fraction of the energy output is converted to electron, rather than to proton, acceleration, the resulting radiation field may inhibit proton acceleration (through energy loss to pion production). The implications of the distance limit to GRBs are discussed in detail in \cite{GRBCRrev}.

The second implication is that for any homogeneous distribution of sources of UHECRs, the drop of $\lambda_E(E_p)$ with increasing $E_p$ will strongly suppress the flux above $\sim0.5\times10^{20}$~eV: At lower energies, sources out to distances $\sim1$~Gpc, comparable to the size of the observable Universe, may contribute to the flux, while at higher energies only local, nearby sources contribute to the flux. This suppression of the flux, usually named the "GZK suppression," is generic to all models where UHECRs are protons produced by a homogeneous distribution of extra-Galactic sources. The question of whether or not such suppression is present in the data is therefore extremely important. The analysis below, aimed at answering this question, follows that of ref.~\cite{BW03}, where interested readers may find more details.

\subsection{Analysis of current data}
\label{sec:data}

In order to address the question of the presence or absence of a GZK suppression, we must adopt a model
for the intrinsic spectrum produced by the UHECR sources (which is subsequently modified by the GZK effect).
We assume that extra-galactic protons in the energy range of $10^{19}$~eV to $10^{21}$~eV are produced by
cosmologically-distributed sources at a rate and spectrum given by
\begin{equation}
E_p^2\frac{d\dot{N}_p}{dE_p}\approx0.6\times 10^{44} {\rm
erg~Mpc^{-3}~yr^{-1}}\phi(z). \label{eq:energyrate}
\end{equation}
An energy spectrum similar to the assumed $dN/dE_p\propto
E_p^{-2}$ has been observed for both
non-relativistic~\cite{CR_data_rev} and
relativistic~\cite{relativistic} shocks. It is believed to to be
due to Fermi acceleration in collisionless
shocks~\cite{CR_data_rev,relativistic}, although a first
principles understanding of the process is not yet available. The
normalization of $0.6\times 10^{44} {\rm erg~Mpc^{-3}~yr^{-1}}$ is
chosen to account for the observed flux. $\phi(z)$ accounts for
redshift evolution ($\phi(z=0)=1$). Motivated by the GRB model
(\S~\ref{sec:GRB}), we have assumed that $\phi(z)$ follows the
evolution of the star-formation rate. The spectrum above
$10^{19}$~eV is only weakly dependent on $\phi(z)$ since proton
energy loss limits their propagation distance. For the heavy
nuclei component dominating at lower, $<10^{19}$~eV, energy we
take the Fly's Eye experimental fit~\cite{fly},
\begin{equation}
\frac{dN}{dE} ~\propto~ E^{-3.50}. \label{eq:galacticspectrum}
\end{equation}
\begin{figure}[htbp]
\epsfxsize=8cm
\centerline{\epsfbox{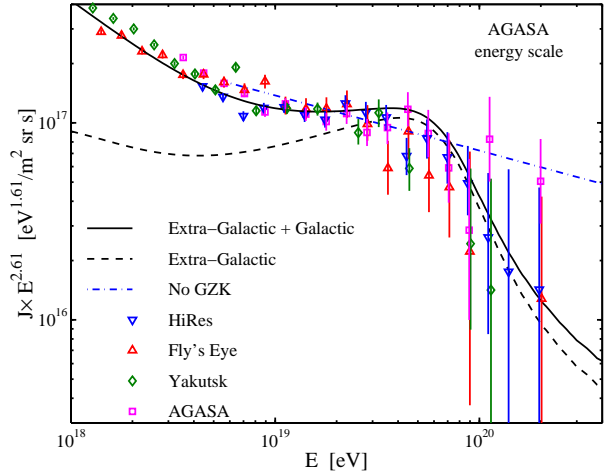}}
\caption{Model versus data. The solid curve shows
the energy spectrum derived from the two-component model discussed in \S~3.2. The dashed
curve shows the extra-Galactic component contribution. The "No GZK" curve is an extrapolation of the
$E^{-2.75}$ energy spectrum derived for the energy range of $6\times 10^{18}$~eV to $4\times 10^{19}$~eV
[1]). Data taken from [11]. AGASA's energy scale was chosen.}
\label{fig:fig_fly_norm}
\end{figure}

Figure~\ref{fig:fig_fly_norm} demonstrates that model predictions are in good agreement with the data of all experiments in the energy range $10^{19}$~eV to $10^{20}$~eV\footnote{As explained in detail in \cite{BW03}, the various experiments are consistent with each other when systematic errors in the absolute energy scale of the events are taken into account. The relative systematic shifts in absolute energy calibration between Fly' Eye and the other experiments, required to bring into agreement the fluxes measured at $10^{19}$~eV by the different experiments, are \{-11\%, +7.5\%, -19\%\} for \{AGASA, HiRes, Yakutsk\}. All shifts are well within the published systematic errors.}. Above $10^{20}$~eV, the Fly's Eye, HiRes and Yakutsk experiments are in agreement with each other and the model. However, the eight AGASA events with energies greater than $10^{20}$~eV disagree with the prediction of the cosmological model. The Fly's Eye, Yakutsk and HiRes experiments have a combined exposure three times that of the AGASA experiment.  The exposures above $10^{20}$~eV are, in units of $10^3{\rm km^2-yr-sr}$: AGASA (1.3), Fly's Eye (0.9), Yakutsk (0.9), and HiRes (2.2). Together, Fly's Eye, Yakutsk, and Hi-Res observe a total of 6 events above $10^{20}$ eV (4 events if the Fly's Eye energy scale is chosen).

How significant is the flux suppression observed by Fly's Eye, HiRes and Yakutsk? The differential energy spectrum observed by the various experiments at the energy range of
$4\times 10^{17}$~eV to $4\times 10^{19}$~eV can be fitted by a broken power-law, where the shallower
component dominating above $\sim6\times 10^{18}$~eV satisfies $J\propto E^{-2.75\pm0.2}$. The number of events observed beyond $10^{20}$ eV by Fly's Eye, HiRes and Yakutsk is at a $>5\sigma$ deficit relative to the number expected from extrapolation to high energy of the low-energy distribution. Adopting the steepest allowed slope, $J\propto E^{-2.95}$, the observed number of events is at a $>3.7\sigma$ deficit.

What is the reason for the discrepancy between AGASA and the other experiments above $10^{20}$~eV? Our analysis shows that the discrepancy is statistically significant. Thus, the discrepancy is most likely due to systematic errors in the estimates of event energies. The Auger detector currently under construction \cite{Auger} is likely to resolve this issue by combining two different detection methods (ground array and nitrogen fluorescence).

\section{GRBs, UHECRs and neutrinos}
\label{sec:GRB}

\begin{figure}[htbp]
\epsfxsize=10cm \centerline{\epsfbox{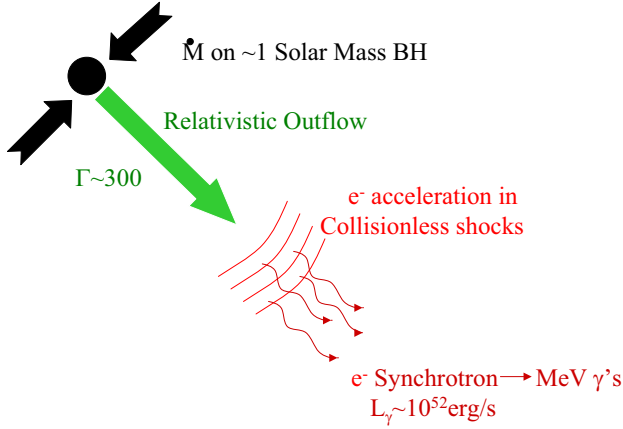}} \caption{The
fireball scenario of GRB production.} \label{fig:GRB}
\end{figure}
Gamma-ray bursts (GRBs) are short, typically tens of seconds long,
flashes of gamma-rays, carrying most of their energy in $>1$~MeV
photons. The detection in the past few years of "afterglows",
delayed X-ray, optical and radio emission from GRB sources, proved
that the sources lie at cosmological distances, and provided
strong support for the scenario of GRB production described in
fig.~\ref{fig:GRB} \cite{GRBrev}. The energy source is believed to
be rapid mass accretion on a newly formed solar-mass black hole
(or, possibly, neutron star). Recent observations suggest that the
formation of the central compact object is associated with type
Ib/c supernovae \cite{GRBSNe}.

The energy release drives an ultra-relativistic,
$\Gamma\sim10^{2.5}$, plasma outflow. At a large distance from the
central black-hole, internal collisionless shocks within the wind,
which arise due to variability in the wind
emitted from the central "engine", accelerate electrons to high
energy. Synchrotron emission from these shock accelerated
electrons is believed to account for the observed $\gamma$-rays.
At still larger distances, the wind impacts on surrounding medium.
Here too, the collisionless shock driven into the ambient gas
accelerates electrons, leading to synchrotron emission which accounts for the "afterglow".

\subsection{The association of GRBs and UHECRs}
\label{sec:GRBCR}

 GRBs were suggested to be UHECR sources in
\cite{W95,VMU}. The GRB-UHECR association was based in \cite{W95}
on two major arguments: (i) The constraints imposed on the
relativistic wind by requiring that it would produce the observed
$\gamma$-rays are remarkably similar to those imposed by the
requirement that the wind would allow proton acceleration to
$10^{20}$~eV; (ii) The rate (per unit volume) at which energy is
generated by GRBs in $\gamma$-rays is similar to the rate at which
energy should be generated in high energy protons in order to
account for the observed UHECR flux.

The constraints on wind parameters are summarized in
table~\ref{table} (for a pedagogical discussion see
\cite{GRBCRrev}).
\begin{table}
  \centering
 \begin{tabbing}
   Wind property  and space \= MeV $\gamma$-rays and more space  \= $10^{20}$~eV protons  \kill
   Wind property  \> MeV $\gamma$-rays  \> $10^{20}$~eV protons  \\
   $u_B/u_e$ \> $\ge0.1$ \> $\ge0.02$ \\
   $\Gamma$ \> $\ge 300$ \> $\ge 100$ \\
   $dN/dE$ \> $dN_e/dE_e\propto E_e^{-2}$ \> $dN_p/dE_p\propto E_p^{-2}$ \\
 \end{tabbing}
  \caption{Constraints on GRB wind parameters from photon observations and
  from the requirement for proton acceleration to $10^{20}$~eV.}\label{table}
\end{table}
$\gamma$-ray observations require the ratio of magnetic field and
electron energy densities, $u_B/u_e$, to exceed 0.1 in order to
produce electron synchrotron emission in the MeV range,
$\Gamma>300$ to avoid pair-production by high energy photon which
would make the wind opaque, and $dN_e/dE_e\propto E_e^{-2}$ to
account for the synchrotron spectrum. Acceleration of protons to
$10^{20}$~eV require $u_B/u_e>0.02$ to satisfy Eq.~\ref{eq:BR} (We
have seen that this equation implies a lower limit to the
luminosity carried by magnetic field, eq.~\ref{eq:L}, and since
the observed photon luminosity represents the wind luminosity
carried by electrons, a lower limit is inferred for the ratio of
magnetic and electron luminosity); $\Gamma>100$ to avoid
synchrotron losses of protons on a time scale shorter than the
acceleration time, and $dN_p/dE_p\propto E_p^{-2}$ to reproduce
the observed UHECR spectrum (see eq.~\ref{eq:energyrate}).

The fact that similar constraints are obtained from independent
arguments, suggests that GRBs and UHECRs are associated. The
association is further supported by the similarity of energy
generation rates. We have shown in \S~\ref{sec:GZK} that the
energy generation rate of protons inferred from UHECR observations
is given by eq.~\ref{eq:energyrate}. The energy generation rate of
GRBs may be estimated as follows. Assuming that the GRB rate
follows the star-formation rate, which is reasonable given the
association with type Ib/c supernovae, the local ($z=0$) GRB rate
is $\approx0.5/{\rm Gpc^3 yr}$ \cite{Schmidt01}. The average MeV
$\gamma$-ray energy release is $\approx3\times10^{53}$~erg for GRBs with known redshift
\cite{Bloom03}. The rate of bursts for which redshift is obtained is smaller than the total observed rate by a factor of $\approx0.7\times0.5=0.35$. The factor $0.7$ is due to the fact that the detection threshold of the BeppoSAX detector that allowed afterglow detection is higher (by a factor $\approx2$) than the threshold of BATSE \cite{Band}, the detector that provided the data based on which the total rate estimate was derived. The factor of $0.5$ is due to the fact that the fraction of bursts for which optical afterglow was detected is $\sim0.5$. Thus, the local ($z=0$) energy generation rate of GRBs in MeV photons is 
\begin{equation}\label{eq:grbrate}
    \dot{\varepsilon}_{\gamma\rm [MeV]}\ge0.35\times 3\times10^{53}{\rm erg} \times 0.5/{\rm Gpc^3 yr}=0.5\times10^{44}{\rm erg/Mpc^3 yr}.
\end{equation} 
This rate is similar to the energy generation rate in UHECRs, eq.~\ref{eq:energyrate}.

\subsection{Predictions: High energy neutrinos}
\label{sec:pred}

Protons accelerated in the fireball to high energy lose energy
through photo-meson interaction with fireball photons. The decay
of charged pions produced in this interaction results in the
production of high energy neutrinos. The key relation is between
the observed photon energy, $E_\gamma$, and the accelerated
proton's energy, $E_p$, at the threshold of the
$\Delta$-resonance. In the observer frame,
\begin{equation}
E_\gamma \,E_{p} = 0.2 \, {\rm GeV^2} \, \Gamma^2\,.
\label{eq:keyrelation}
\end{equation}
For $\Gamma\approx300$ and $E_\gamma=1$~MeV, we see that
characteristic proton energies $\sim 10^{16}$~eV are required to
produce pions. Since neutrinos produced by pion decay typically
carry $5\%$ of the proton energy, production of $\sim 10^{14}$~eV
neutrinos is expected \cite{WnB97}.

The fraction of energy lost by protons to pions, $f_\pi$, is $f_\pi\approx0.2$
\cite{GRBCRrev}. Assuming that GRBs generate the observed UHECRs, the expected
GRB muon-neutrino flux may be estimated using eq.~\ref{eq:energyrate} \cite{GRBCRrev}, 
\begin{equation}
E_\nu^2\Phi_{\nu} \approx
\frac{c}{4\pi}\frac{f_\pi}{4}E_p^2 (d\dot N_p/dE_p)t_H \approx
0.3\times10^{-8}{f_\pi\over0.2}{\rm GeV\,cm}^{-2}{\rm s}^{-1}{\rm
sr}^{-1}. \label{eq:JGRB}
\end{equation}
Here $t_H$ is the Hubble time and the factor $1/4$ multiplying $f_\pi$ is due to the fact that in  photo-production of pions charged and neutral pions are created with roughly equal probability, and when a charged pion decays roughly half its energy is carried by muon neutrinos. This neutrino spectrum extends to $\sim10^{16}$~eV, and suppressed at higher energy due to energy loss of pions and muons. 
Eq.~\ref{eq:JGRB} implies a detection rate of $\sim20$ neutrino-induced muon events per year (over $4\pi$~sr) in a cubic-km detector. Since GRB neutrino events are correlated both in time and in direction with gamma-rays, their detection is practically background free.

High energy neutrinos may be produced also in other stages of fireball evolution. For a detailed discussion see \cite{GRBCRrev,RMW03} and references therein.

\section{Giga-ton neutrino detectors and UHECRs}
\label{sec:GTnu}

If UHECRs are protons of extra-Galactic origin then, regardless of the nature of their sources, their \begin{figure}[htbp]
\epsfxsize=10cm
\centerline{\epsfbox{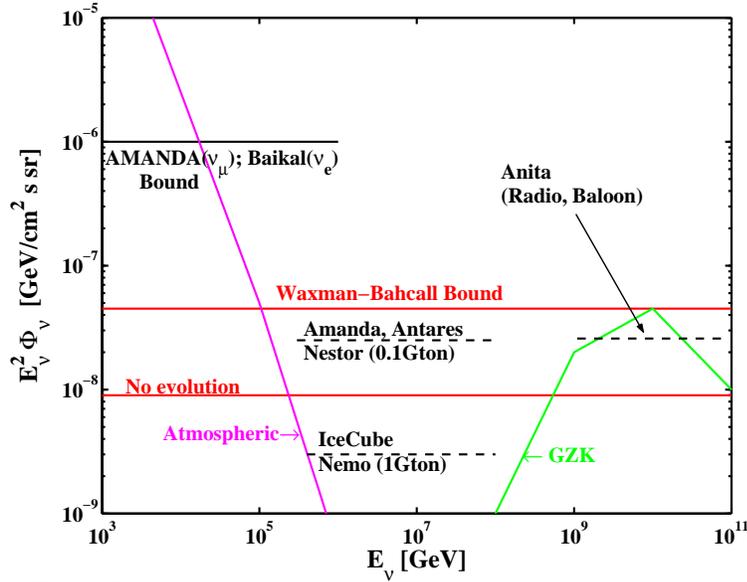}}
\caption{The upper bound imposed by UHECR observations on the extra-Galactic high energy muon neutrino intensity (lower-curve: no evolution of the energy production rate of eq.~\ref{eq:energyrate}, upper curve: assuming evolution following star formation rate), compared with the atmospheric neutrino background and with the experimental upper bound of BAIKAL [23] and AMANDA [24]. The curve labelled "GZK" shows the intensity due to interaction with micro-wave background photons. Dashed curves show the expected sensitivity of 0.1Gton (AMANDA, ANTARES [25], NESTOR [26]) and 1Gton (IceCube [27], NEMO [28]) Cerenkov detectors, and of the balloon radio experiment ANITA [29].}
\label{fig:WBbound}
\end{figure}
existence implies the existence of a flux of extra-Galactic high energy neutrinos. Such neutrinos are expected to be produced by the decay of charged pions produced in interactions of UHECRs with photons (or nucleons). UHECR observations set an upper limit to the intensity of high energy neutrinos produced in sources where the optical depth to nucleon interaction with photons is not high, i.e. where the average number of photo-production interactions the nucleon undergoes before escaping is not high \cite{WBbound}. This upper limit is obtained by taking $f_\pi=1$ in eq.~\ref{eq:JGRB}, i.e. by assuming that all the UHECR energy is converted to pions. The resulting bound, shown in fig.~\ref{fig:WBbound}, applies to GRBs and to the observed jets of AGN, which satisfy the requirement of small photo-production optical depth. Sources with large optical depth may exist, and may therefore produce a flux exceeding the bound. However, we have no direct evidence for the existence of such sources. 

Fig.~\ref{fig:WBbound} demonstrates that the detection (by muon optical Cerenkov detectors) of the expected fluxes in the energy range of 1~TeV to 1000~TeV requires cubic-km (i.e. Gton) scale detectors. The GZK flux (which "touches" the bound at $\sim10^{19}$~eV because ultra-high energy protons produced at large distances lose all their energy to pion production in interaction with microwave background photons) requires still larger detectors (due to the lower number flux at higher energy). For a more general, yet concise, discussion of high energy neutrino telescopes and their science goals see chapters 2 and 3 of \cite{HENAP}. For more detailed review see \cite{HnH02}.

\section{Conclusions}
\label{sec:discussion}

Detectors of high energy cosmic-ray and neutrinos currently under construction may allow to identify the sources of UHECRs. Such identification will provide only a partial resolution of the puzzle. A major challenge will remain in understanding the physics of the sources. GRBs and AGN are the most powerful astronomical objects, and are likely candidates for the production of ultra-high energy protons and neutrinos. In both, the energy source is likely to be mass accretion onto a black hole, leading to relativistic outflows. The models describing these objects are largely phenomenological, and major open questions remain regarding the underlying physics. Data from the new experiments may allow to resolve some of these open questions.

High energy neutrinos are expected to be produced in astrophysical sources by the decay of charged pions, 
which lead to the production of muon and electron neutrinos.
However, if the atmospheric neutrino anomaly has the explanation it is
usually given, oscillation to $\nu_\tau$'s with mass $\sim0.1{\rm\ eV}$
\cite{Osc}, then one should detect equal numbers of $\nu_\mu$'s and $\nu_\tau$'s.
Up-going $\tau$'s, rather than $\mu$'s, would be a distinctive signature of such oscillations.
Since $\nu_\tau$'s are not expected to be produced, looking
for $\tau$'s would be an "appearance experiment."

Detection of neutrinos from GRBs could be used to test the simultaneity of
neutrino and photon arrival to an accuracy of $\sim1{\rm\ s}$, checking the assumption of
special relativity that photons and neutrinos have the same limiting speed.
These observations would also test the weak equivalence principle, according to which photons and neutrinos should suffer the same time delay as they pass through a gravitational potential.
With $1{\rm\ s}$ accuracy, a burst at $1{\rm\ Gpc}$ would reveal
a fractional difference in limiting speed
of $10^{-17}$, and a fractional difference in gravitational time delay
of order $10^{-6}$ (considering the Galactic potential alone).
Previous applications of these ideas to supernova 1987A (see \cite{John_book} for review), yielded much weaker upper limits: of order $10^{-8}$ and $10^{-2}$ respectively.

\end{document}